\documentclass[preprintnumbers, prd, showpacs, floatfix,twocolumn,
preprintnumbers, letterpaper, superscriptaddress,nofootinbib]{revtex4}
\usepackage{amsfonts}
\usepackage{amsmath}
\usepackage{amssymb,epsf}
\usepackage{latexsym}
\usepackage{graphicx,epsfig}
\usepackage{amssymb}
\usepackage{subfigure}
\usepackage{epstopdf}

\begin{document}

\title{Thermodynamics of higher dimensional topological dilaton black
holes with power-law Maxwell field}

\author{M. Kord Zangeneh,$^{1}$\thanks{mkzangeneh@shirazu.ac.ir}, A. Sheykhi$^{1,2}$
\thanks{asheykhi@shirazu.ac.ir}
and M. H. Dehghani \thanks{mhd@shirazu.ac.ir}}
\address{Physics Department and Biruni Observatory, College of
Sciences, Shiraz University, Shiraz 71454, Iran}
        \address{ Research Institute for Astronomy and Astrophysics of Maragha
         (RIAAM), P.O. Box 55134-441, Maragha, Iran}

\begin{abstract}
In this paper, we extend the study on the nonlinear power-law
Maxwell field to dilaton gravity. We introduce the
$(n+1)$-dimensional action in which gravity is coupled to a
dilaton and power-law nonlinear Maxwell field, and obtain the
field equations by varying the action. We construct a new class of
higher dimensional topological black hole solutions of
Einstein-dilaton theory coupled to a power-law nonlinear Maxwell
field and investigate the effects of the nonlinearity of the
Maxwell source as well as the dilaton field on the properties of
the spacetime. Interestingly enough, we find that the solutions
exist provided one assumes three Liouville-type potentials for the
dilaton field, and in case of the Maxwell field one of the
Liouville potential vanishes. After studying the physical
properties of the solutions, we compute the mass, charge, electric
potential and temperature of the topological dilaton black holes.
We also study thermodynamics and thermal stability of the
solutions and disclose the effects of the dilaton field and the
power-law Maxwell field on the thermodynamics of these black
holes. Finally, we comment on the dynamical stability of the
obtained solutions in four-dimensions.
\end{abstract}

\pacs{04.70.Bw, 04.30.-w, 04.70.Dy}

\maketitle

\address{Physics Department and Biruni Observatory, College of
Sciences, Shiraz University, Shiraz 71454, Iran}
\address{ Research Institute for Astronomy and Astrophysics of Maragha
         (RIAAM), P.O. Box 55134-441, Maragha, Iran}

\address{Physics Department and Biruni Observatory, College of
Sciences, Shiraz University, Shiraz 71454, Iran}
\address{ Research Institute for Astronomy and Astrophysics of Maragha
         (RIAAM), P.O. Box 55134-441, Maragha, Iran}

\section{Introduction}

At the present epoch, the Universe expands with acceleration instead of
deceleration along the scheme of standard Friedmann model \cite{Expan}. This
fact created much more interest in the alternative theories of gravity in
recent years. One of the alternative theories of gravity is dilatn gravity
which can be thought as the low energy limit of string theory. Indeed, in
the low energy limit of string theory, one recovers Einstein gravity along
with a scalar dilaton field which is nonminimally coupled to the gravity and
other fields such as gauge fields \cite{Wit1}. The action of dilaton gravity
also contains one or more Liouville-type potentials, which can be resulted
by the breaking of spacetime supersymmetry in ten dimensions.

Many attempts have been made to construct exact solutions of
Einstein-Maxwell-dilaton (EMd) gravity in the literature. For instance,
exact asymptotically flat solutions of EMd gravity with no dilaton potential
have been constructed in Refs. \cite{CDB1,CDB2,MW,PW}. But, the asymptotic
behavior of the solutions of dilaton gravity with potential may be neither
flat nor (anti)-de Sitter [(A)dS]. These kind of solutions which are neither
asymptotically flat nor (A)dS are interesting from different points of view.
First, it is speculated that the linear dilaton spacetimes which appear as
near-horizon limits of the dilatonic black holes may show holography that
can be considered as an indication of the possible extensions of AdS/CFT
correspondence \cite{Ahar}. Second, the range of validity of methods and
tools originally developed for, and tested in the case of, asymptotically
flat or asymptotically AdS black holes may be extended using such solutions.
Third, in addition to black holes with spherical horizon, there exist black
hole solutions with toroidal or hyperbolic event horizons, as in the case of
asymptotically AdS solutions. Having different topologies for the horizon
gives rise to different properties for the black holes with drastically
different black holes thermodynamics. For instance, it was argued that
Schwarzschild-AdS black holes with toroidal or hyperbolic horizons are
thermally stable and the Hawking-Page phase transition \cite{Haw3} does not
occur \cite{Birm}, while Schwarzschild black holes with spherical horizon
are not stable. The topological black holes are studied extensively in many
aspects \cite{Shey2,Lemos,Cai2,Bril1,Cai3,Cai4,Cri,MHD,other,Ban}. Many
authors have been explored asymptotically non-flat or non-(A)dS black hole
solutions \cite{Shey2,MW,PW,CHM,Cai,Clem,Mitra,Shey0,
SR,DF,Dehmag,SDR,SDRP,yaz,yaz2,SRM}. Static charged black hole solutions in
the presence of Liouville-type potential, with positive \cite{CHM}, zero or
negative constant curvature horizons \cite{Cai} have been discovered and
properties of these solutions which are not asymptotically (A)dS have been
studied \cite{Clem}. Also, thermodynamics of $(n+1)$-dimensional dilaton
black holes with unusual asymptotics have been studied \cite{Shey,DHSR}.

Here, we turn the investigations on the dilaton gravity to includes the
power-law Maxwell field term in the action. This is motivated by the fact
that, as in the case of scalar field which has been shown that particular
power of the massless Klein-Gordon Lagrangian shows conformal invariance in
arbitrary dimension \cite{Haissene}, one can have a conformally
electrodynamic Lagrangian in higher dimensions. Although Maxwell Lagrangian
loses its conformally invariant property in higher dimensions, the
Lagrangian $[-\exp (-4\alpha \Phi /(n-1))F_{\mu \nu }F^{\mu \nu }]^{(n+1)/4}$
is conformally invariant in $(n+1)$ dimensions. That is, this Lagrangian is
invariant under the conformal transformation $g_{\mu \nu }\rightarrow \Omega
^{2}g_{\mu \nu }$ and $A_{\mu }\rightarrow A_{\mu }$. The studies on the
black object solutions coupled to a conformally invariant Maxwell field were
carried out in \cite{Hass1,confshey2}.

The motivation of studying solutions of Einstein gravity with arbitrary
dimensions is based on the string theory which predicts spacetime to have
more than four dimensions. Although it was a thought for a while that the
extra spacial dimensions are of the order of Planck scale, recent theories
suggest that if we live on a $3$-dimensional brane in a higher dimensional
bulk it is possible to have the extra dimensions relatively large and still
unobservable \cite{RS,DGP}. All gravitational objects including black holes
are higher dimensional in such a scenario. Higher dimensional Ricci flat
rotating black branes with a conformally invariant power-Maxwell source in
the absence of a dilaton field have been investigated in \cite{hendisedehi}.
Thermodynamics of higher dimensional topological dilaton black holes with
linear Maxwell source have been explored in \cite{Shey2}.

In this paper, we would like to construct a new class of $(n+1)$-dimensional
topological black holes of dilaton gravity in the presence of power-law
Maxwell field $[-\exp (-4\alpha \Phi /(n-1))F_{\mu \nu }F^{\mu \nu }]^{p}$,
where we relax the conformally invariant issue for generality. Of course,
the solution exists for the case of conformally invariant source $p=(n+1)/4$%
. We find that the solution exists provided one assumes three Liouville-type
potentials. The interesting point is that one of the Liouville potentials
vanish for the case of Maxwell field ($p=1$). We shall investigate the
thermal stability of the black holes and explore the effects of nonlinearity
of Maxwell field on the thermodynamics of these black holes.

This paper is structured as follows. In Sec. \ref{Field}, we introduce the
action of Einstein-dilaton gravity coupled to power-law Maxwell field, and
by varying the action we obtain the field equations. Then, we construct the
exact topological black hole solutions of this theory and investigate their
properties. In Sec. \ref{Therm}, we obtain the conserved and thermodynamic
quantities of the solutions and verify the validity of the first law of
black hole thermodynamics. In Sec. \ref{Stab}, we study thermal stability of
the solutions in both canonical and grand canonical ensembles. The last
section is devoted to conclusions and discussions.

\section{Field Equations and Solutions\label{Field}}

The action of $(n+1)$-dimensional $(n\geq 3)$ Einstein-power Maxwell-dilaton
gravity can be written as%
\begin{eqnarray}
S &=&\frac{1}{16\pi }\int d^{n+1}x\sqrt{-g}\left\{ \mathcal{R}-\frac{4}{n-1}%
(\nabla \Phi )^{2}\right.  \notag \\
&&\left. -V(\Phi )+\left( -e^{-4\alpha \Phi /(n-1)}F\right) ^{p}\right\} ,
\label{Act}
\end{eqnarray}%
where $\mathcal{R}$ is the Ricci scalar, $\Phi $ is the dilaton field,$%
V(\Phi )$ is a potential for $\Phi $, and $p$ and $\alpha $ are two
constants determining the nonlinearity of the electromagnetic field and the
strength of coupling of the scalar and electromagnetic field, respectively. $%
F=F_{\lambda \eta }F^{\lambda \eta }$, where $F_{\mu \nu }=\partial _{\mu
}A_{\nu }-\partial _{\nu }A_{\mu }$ is the electromagnetic field tensor and $%
A_{\mu }$ is the electromagnetic potential. The equations of motion can be
obtained by varying the action (\ref{Act}) with respect to the gravitational
field $g_{\mu \nu }$, the dilaton field $\Phi $ and the gauge field $A_{\mu
} $ which yields the following field equations
\begin{eqnarray}
&&\mathcal{R}_{\mu \nu }=g_{\mu \nu }\left\{ \frac{1}{n-1}V(\Phi )+\frac{%
(2p-1)}{n-1}\left( -Fe^{-4\alpha \Phi /(n-1)}\right) ^{p}\right\}  \notag \\
&&+\frac{4}{n-1}\partial _{\mu }\Phi \partial _{\nu }\Phi +2pe^{-4\alpha
p\Phi /(n-1)}(-F)^{p-1}F_{\mu \lambda }F_{\nu }^{\text{ \ }\lambda },
\label{FE1}
\end{eqnarray}

\begin{eqnarray}
\nabla ^{2}\Phi -\frac{n-1}{8}\frac{\partial V}{\partial \Phi }-\frac{%
p\alpha }{2}e^{-{4\alpha p\Phi }/({n-1})}(-F)^{p} &=&0,  \label{FE2} \\
\partial _{\mu }\left( \sqrt{-g}e^{-{4\alpha p\Phi }/({n-1}%
)}(-F)^{p-1}F^{\mu \nu }\right) &=&0.  \label{FE3}
\end{eqnarray}%
We like to find the static topological solutions of the above field
equations. The most general form of such a metric can be written as
\begin{equation}
ds^{2}=-f(r)dt^{2}+{\frac{dr^{2}}{f(r)}}+r^{2}R^{2}(r)h_{ij}dx^{i}dx^{j},
\label{metric}
\end{equation}%
where $f(r)$ and $R(r)$ are functions of $r$ which should be determined, and
$h_{ij}$ is a function of coordinates $x^{i}$ which spanned an $(n-1)$%
-dimensional hypersurface with constant scalar curvature $(n-1)(n-2)k$. Here
$k$ is a constant and characterizes the hypersurface. Without loss of
generality, one can take $k=0,1,-1$, such that the black hole horizon in (%
\ref{metric}) can be a zero (flat), positive (spherical) or negative
(hyperbolic) constant curvature hypersurface. The Maxwell equation (\ref{FE3}%
) can be integrated immediately to give
\begin{equation}
F_{tr}=\frac{qe{^{{\frac{4\alpha \,p\Phi \left( r\right) }{\left( n-1\right)
\left( 2\,p-1\right) }}}}}{\left( rR\right) ^{{\frac{n-1}{2\,p-1}}}},
\label{Ftr}
\end{equation}%
where $q$ is an integration constant related to the electric charge of the
black hole. Substituting (\ref{metric}) and (\ref{Ftr}) in the field
equations (\ref{FE1}) and (\ref{FE2}), we arrive at

\begin{gather}
f^{\prime \prime }+\,{\frac{\left( n-1\right) f^{\prime }}{r}}+{\frac{\left(
n-1\right) f^{\prime }R^{\prime }}{R}}+{\frac{2V}{n-1}}  \notag \\
-\frac{2[1+\left( n-3\right) ]p}{n-1}\left( 2\,{q}^{2}\left( rR\right) ^{-{%
\frac{2(n-1)}{2\,p-1}}}e{^{\,{\frac{4\alpha \,\Phi }{\left( n-1\right)
\left( 2\,p-1\right) }}}}\right) ^{p}=0,  \label{fe1}
\end{gather}

\begin{gather}
f^{\prime \prime }+{\frac{\left( n-1\right) f^{\prime }}{r}}+{\frac{\left(
n-1\right) f^{\prime }R^{\prime }}{R}}+{\frac{2V}{n-1}}  \notag \\
+{\frac{4\left( n-1\right) fR^{\prime }}{rR}}+{\frac{2\left( n-1\right)
fR^{\prime \prime }}{R}}+\,{\frac{8f\Phi ^{\prime 2}}{n-1}}  \notag \\
-\frac{2[1+\left( n-3\right) p]}{n-1}\left( 2\,{q}^{2}\left( rR\right) ^{-{%
\frac{2(n-1)}{2\,p-1}}}e{^{\,{\frac{4\alpha \,\Phi }{\left( n-1\right)
\left( 2\,p-1\right) }}}}\right) ^{p}=0,  \label{fe2}
\end{gather}

\begin{gather}
{\frac{f^{\prime }}{r}}+{\frac{f^{\prime }R^{\prime }}{R}+\frac{\left(
n-2\right) f}{{r}^{2}}+\frac{2\left( n-1\right) fR^{\prime }}{rR}}  \notag \\
{+\frac{\left( n-2\right) R^{\prime 2}f}{R^{2}}+\frac{fR^{\prime \prime }}{R}%
}-{\frac{k\left( n-2\right) }{\left( rR\right) ^{2}}+\frac{V}{n-1}}  \notag
\\
+\frac{2\,p-1}{n-1}\left( 2\,{q}^{2}\left( rR\right) ^{-{\frac{2(n-1)}{2\,p-1%
}}}e{^{\,{\frac{4\alpha \,\Phi }{\left( n-1\right) \left( 2\,p-1\right) }}}}%
\right) ^{p}=0,  \label{fe3}
\end{gather}

\begin{gather}
f\Phi ^{\prime \prime }+\Phi ^{\prime }f^{\prime }+{\frac{\left( n-1\right)
f\Phi ^{\prime }}{r}}  \notag \\
+{\frac{\left( n-1\right) f\Phi ^{\prime }R^{\prime }}{R}}-\frac{n-1}{8}%
\frac{dV}{d\Phi }  \notag \\
-\frac{\,p\alpha }{2}\,\left( 2\,{q}^{2}\left( rR\right) ^{-{\frac{2(n-1)}{%
2\,p-1}}}e{^{\,{\frac{4\alpha \,\Phi }{\left( n-1\right) \left(
2\,p-1\right) }}}}\right) ^{p}=0,  \label{fe4}
\end{gather}%
where the prime denotes derivative with respect to $r$. Our aim here is to
construct exact, $(n+1)$-dimensional topological solutions of the above
field equations with an arbitrary dilaton coupling parameter $\alpha $.
Calculations show that there exist exact topological solutions of physically
interest provided we take the dilaton potential with three Liouville-type
potentials as
\begin{equation}
V(\Phi )=2\,\Lambda _{1}e{^{2\zeta _{1}\,\Phi }}+2\Lambda _{2}e{^{2\zeta
_{2}\,\Phi }}+2\,\Lambda e{^{2\zeta _{3}\,\Phi }},  \label{v2}
\end{equation}%
where $\Lambda _{1}$, $\Lambda _{2}$, $\Lambda $, $\zeta _{1}$, $\zeta _{2}$
and $\zeta _{3}$ are constants. It is important to note that in case of
topological black holes of EMd theory, one only needs to take two terms in
the Liouville potential \cite{Shey2}, while here we find that for power-law
Maxwell source in dilaton gravity we need to add an additional term to the
potential and consider Liouville-type dilaton potential with three terms.

In order to solve the system of equations (\ref{fe1})-(\ref{fe4}) for three
unknown functions $f(r)$, $R(r)$ and $\Phi (r)$, we make the ansatz
\begin{equation}
R(r)=e{^{{{2\alpha \,\Phi \left( r\right) }/({n-1})}}}.  \label{Rphi}
\end{equation}%
Subtracting (\ref{fe1}) from (\ref{fe2}), after using (\ref{Rphi}), we find
\begin{equation}
\Phi ^{\prime \prime }+\frac{2(\alpha ^{2}+1)\Phi ^{\prime 2}}{\alpha (n-1)}+%
\frac{2\Phi ^{\prime }}{r}=0,
\end{equation}%
which has the following solution
\begin{equation}
\Phi (r)=\frac{\,\left( n-1\right) \alpha }{2\left( {\alpha }^{2}+1\right) }%
\,\ln \left( {\frac{b}{r}}\right) .  \label{phi}
\end{equation}%
Substituting (\ref{Rphi}) and (\ref{phi}) in Eqs. (\ref{fe2})-(\ref{fe4}),
one can easily show that these equation have a unique consistent solution of
the form
\begin{eqnarray}
f(r) &=&\frac{k\left( n-2\right) (1+\alpha ^{2})^{2}{r}^{2\gamma }}{%
(1-\alpha ^{2})\left( {\alpha }^{2}+n-2\right) {b}^{2\gamma }}-\frac{m}{{r}%
^{(n-1)(1-\gamma )-1}}  \notag  \label{f} \\
&&+\frac{2^{p}p(1+\alpha ^{2})^{2}\left( 2p-1\right) ^{2}{b}^{-{\frac{%
2\left( n-2\right) p\gamma }{\left( 2\,p-1\right) }}}{q}^{2\,p}\,}{\Pi
\left( n+{\alpha }^{2}-2\,p\right) {r}^{-\frac{2[\,\left( n-3\right)
p+1]-2p\left( n-2\right) \gamma }{2p-1}}}  \notag \\
&&-\frac{2\Lambda {b}^{2\gamma }(1+\alpha ^{2})^{2}{r}^{2(1-\gamma )}}{%
\left( n-1\right) \left( n-{\alpha }^{2}\right) },  \label{fr}
\end{eqnarray}%
where $b$ is an arbitrary non-zero positive constant, $\gamma =\alpha
^{2}/(\alpha ^{2}+1)$, $\Pi ={\alpha }^{2}+\left( n-1-{\alpha }^{2}\right) p$%
, and the constants should be fixed as
\begin{gather}
\zeta _{1}={\frac{2}{\left( n-1\right) \alpha }},\hspace{0.8cm}\zeta _{2}={%
\frac{2p\left( n-1+{\alpha }^{2}\right) }{\left( n-1\right) \left(
2\,p-1\right) \alpha }},  \notag \\
\zeta _{3}=\,{\frac{2\alpha }{n-1}},\hspace{0.8cm}\Lambda _{1}={\frac{%
k\left( n-1\right) \left( n-2\right) {\alpha }^{2}}{2{b}^{2}\left( {\alpha }%
^{2}-1\right) }},  \notag \\
\Lambda _{2}=\frac{2^{p-1}\left( 2\,p-1\right) \left( p-1\right) {\alpha }%
^{2}\,{q}^{2\,p}}{\Pi {b}^{{\frac{2\left( n-1\right) p}{2\,p-1}}}}.
\label{lam0}
\end{gather}%
It is worth noting that in the linear Maxwell case where $p=1$, we have $%
\Lambda _{2}=0$ and hence the potential has two terms. Indeed, the term $%
2\Lambda _{2}e{^{2\zeta _{2}\,\Phi }}$ in the Liouville potential is
necessary in order to have solution (\ref{fr}) for the field equations of
power-law Maxwell field in dilaton gravity. Note that $\Lambda $ remains as
a free parameter which plays the role of the cosmological constant and we
assume to be negative and take it in the standard form $\Lambda
=-n(n-1)/2l^{2}$. The parameter $m$ in Eq. (\ref{fr}) is the integration
constant which is known as the geometrical mass and can be written in term
of horizon radius as%
\begin{eqnarray}
m(r_{+}) &=&\frac{k\left( n-2\right) {b}^{-2\gamma }{r}_{+}^{{\frac{{\alpha }%
^{2}+n-2}{{\alpha }^{2}+1}}}}{(2\gamma -1)(\gamma -1)\left( {\alpha }%
^{2}+n-2\right) }  \notag \\
&&+\frac{2^{p}p\,\left( 2p-1\right) ^{2}{b}^{{\frac{-2\,\left( n-2\right)
\gamma p}{\left( 2\,p-1\right) }}}{q}^{2\,p}{r}_{+}^{-{\frac{{\alpha }%
^{2}-2\,p+n}{\left( 2\,p-1\right) \left( {\alpha }^{2}+1\right) }}}}{(\gamma
-1)^{2}\left( {\alpha }^{2}-2p+n\right) \Pi }\,  \notag \\
&&+\frac{{b}^{2\gamma \,}n{r}_{+}^{-{\frac{{\alpha }^{2}-n}{{\alpha }^{2}+1}}%
}}{l^{2}(\gamma -1)^{2}\left( n-{\alpha }^{2}\right) },  \label{mrh}
\end{eqnarray}%
where $r_{+}$ is the positive real root of $f(r_{+})=0$. In the limiting
case where $p=1$, solution (\ref{fr}) reduces to the topological dilaton
black holes of EMd gravity presented in Ref. \cite{Shey2}. One may note that
in the absence of a non-trivial dilaton ($\alpha =\gamma =0$) for a linear
Maxwell theory ($p=1$), solution (\ref{fr}) reduces to%
\begin{equation}
f(r)=k-\frac{m}{r^{n-2}}+\frac{2q^{2}}{(n-1)(n-2)r^{2(n-2)}}-\frac{2\Lambda
}{n(n-1)}r^{2},
\end{equation}%
which describes an $(n+1)$-dimensional asymptotically AdS topological black
hole with a positive, zero or negative constant curvature hypersurface (see
for example \cite{Bril1,Cai3}). One can easily show that the gauge potential
$A_{t}$ corresponding to the electromagnetic field (\ref{Ftr}) can be
written as
\begin{equation}
A_{t}=\frac{q{b}^{{\frac{\left( 2\,p+1-n\right) \gamma }{\left(
2\,p-1\right) }}}}{\Upsilon {r}^{\Upsilon }},  \label{At}
\end{equation}%
where $\Upsilon ={(n-2p+\alpha }^{2}{)/[(2p-1)(1+\alpha }^{2})]$. Let us
discuss the range of parameters $p$ and $\alpha $ for which our obtained
solutions have reasonable behavior and physically are more interesting for
us. There are two restrictions on $p$ and $\alpha $. The first one is due to
the fact that the electric potential $A_{t}$ should have a finite value at
infinity. This leads to $\Upsilon >0$:
\begin{equation}
\frac{{n-2p+}\alpha ^{2}}{({2p-1)(1+}\alpha ^{2})}>0.  \label{res2}
\end{equation}%
\ \ \ \ The above equation leads to the following restriction on the range
of $p$%
\begin{equation}
\frac{1}{2}<p<\frac{n+\alpha ^{2}}{2}.  \label{res3}
\end{equation}%
The second restriction comes from the fact that the term including $m$ in
spacial infinity should vanish. This fact leads to the following restriction
on $\alpha $
\begin{equation}
\alpha ^{2}<n-2.  \label{res5}
\end{equation}%
Thus, one can summarize (\ref{res3}) and (\ref{res5}) as follows:
\begin{eqnarray}
\text{For }\frac{1}{2} &<&p<\frac{n}{2}\text{, \ \ \ \ \ \ \ \ }0\leq \alpha
^{2}<n-2,  \label{res6} \\
\text{For }\frac{n}{2} &<&p<n-1\text{, \ \ \ \ }2p-n<\alpha ^{2}<n-2.
\label{res7}
\end{eqnarray}%
It is worth mentioning that in the above ranges the dilaton potential $%
V(\Phi )$ has a lower finite limit in the range $1/2<p<1$, where $\Lambda
_{2}<0$ and therefore the system is stable. Also one can easily see that in
the above ranges, as in the special cases of $p=1$ or $\alpha =0$, the term
including $q$ in $f(r)$ vanishes at spacial infinity as one expects. Also,
one may note that in the allowed ranges of $p$ and $\alpha $, $\Pi $ is
always positive and therefore the $q$ term in $f(r)$ is always positive.

Next, we study the physical properties of the solutions. First, we
investigate the asymptotic behavior of the solutions. For $\alpha <1$, the
first term in $f(r)$ is dominant at infinity. Thus, in order to have a
positive value for $f(r)$ at infinity, $\Lambda <0$. On the other hand for $%
\alpha >1$, the second term in the metric function is dominant at large $r$
and therefore $k$ should be equal to $-1$ or zero. It is notable to mention
that these solutions do not exist for the string case where $\alpha =1$ in
the $k=\pm 1$ cases. Also, it is worth mentioning that the solution is
well-defined in the allowed ranges of $\alpha $\ and $p$. Next, we look for
the curvature singularities. The Kretschmann scalar $R_{\mu \nu \lambda
\kappa }R^{\mu \nu \lambda \kappa }$ diverges at $r=0$, it is finite for $%
r\neq 0$ and goes to zero as $r\rightarrow \infty $. Thus, there is an
essential singularity located at $r=0$.

As we mentioned, the charge term is positive every where and since the
dominant term is the charge term as $r$ goes to zero, the singularity is
timelike as in the case of Reissner-Nordstrom black holes. Thus, one cannot
have a Schwarzschild-type black hole solution with one event horizon. In
order to consider the type of singularity whether it is naked or not, we
calculate the Hawking temperature of the topological black holes. The
Hawking temperature can be written as

\begin{eqnarray}
T_{+} &=&\frac{f^{\prime }(r_{+})}{4\pi }=\frac{(1+\alpha ^{2})}{4\pi }%
\left\{ \frac{k\left( n-2\right) }{b^{2\gamma }(1-\alpha
^{2})r_{+}^{1-2\gamma }}\right.  \notag  \label{Th} \\
&&\left. -\frac{\Lambda b^{2\gamma }r_{+}^{1-2\gamma }}{n-1}-\frac{%
2^{p}p\left( 2p-1\right) {b}^{\,{\frac{-2\left( n-2\right) \gamma p}{\left(
2\,p-1\right) }}}{q}^{2\,p}}{\Pi {r}_{+}^{\frac{2p(n-2)(1-\gamma )+1}{2p-1}}}%
\right\} .  \notag
\end{eqnarray}%
Extreme black holes occurs when $r_{+}$ or $q$ are chosen such that $T_{+}=0$%
. Using (\ref{Th}), one can find that

\begin{eqnarray}
q_{\mathrm{ext}}^{2p} &=&\frac{{b}^{{\frac{2p\left( n-2\right) \gamma }{%
\left( 2\,p-1\right) }}}\Pi }{{p}\left( 2\,p-1\right) 2^{p}}{r}_{\mathrm{ext}%
}^{\frac{2p(n(1-\gamma )-1)+2\gamma }{2p-1}}  \notag \\
&&\times \left[ \frac{n{b}^{2\gamma }}{l^{2}}+\frac{k\left( n-2\right) }{%
\left( 1-{\alpha }^{2}\right) {b}^{2\gamma }}{r}_{\mathrm{ext}}^{{4\gamma -2}%
}\right] ,
\end{eqnarray}%
which for the case of $\alpha =0$, reduces to

\begin{equation}
\left( q_{\mathrm{ext}}^{2p}\right) _{\alpha =0}=\frac{p\left( n-1\right) {r}%
_{\mathrm{ext}}^{\frac{2(n-1)p}{2p-1}}}{2^{p}{p}\left( 2\,p-1\right) }\left(
\frac{n}{l^{2}}+\frac{k\left( n-2\right) }{{r}_{\mathrm{ext}}^{{2}}}\right) .
\notag
\end{equation}%
Thus, our solutions present black holes with inner and outer horizons
located at $r_{-}$ and $r_{+}$ provided $q<q_{\mathrm{ext}}$, an extreme
black hole if $q=q_{\mathrm{ext}}$ and a naked singularity provided $q>q_{%
\mathrm{ext}}$ (see figs. \ref{fig6}-\ref{fig8}).

\begin{figure}[tbp]
\epsfxsize=7cm \centerline{\epsffile{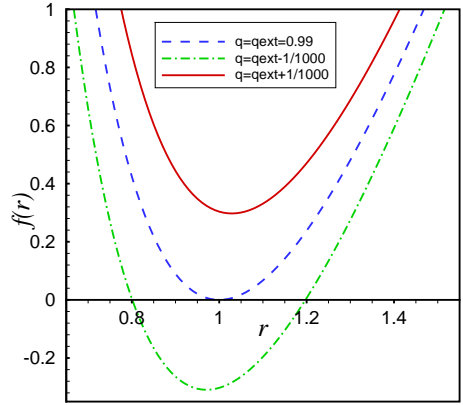}}
\caption{The function $f(r)$ versus $r$ for $n=4$, $\protect\alpha =0.5$, $%
p=2$, $l=b=1$ and $k=0$. }
\label{fig1}
\end{figure}

\begin{figure}[tbp]
\epsfxsize=7cm \centerline{\epsffile{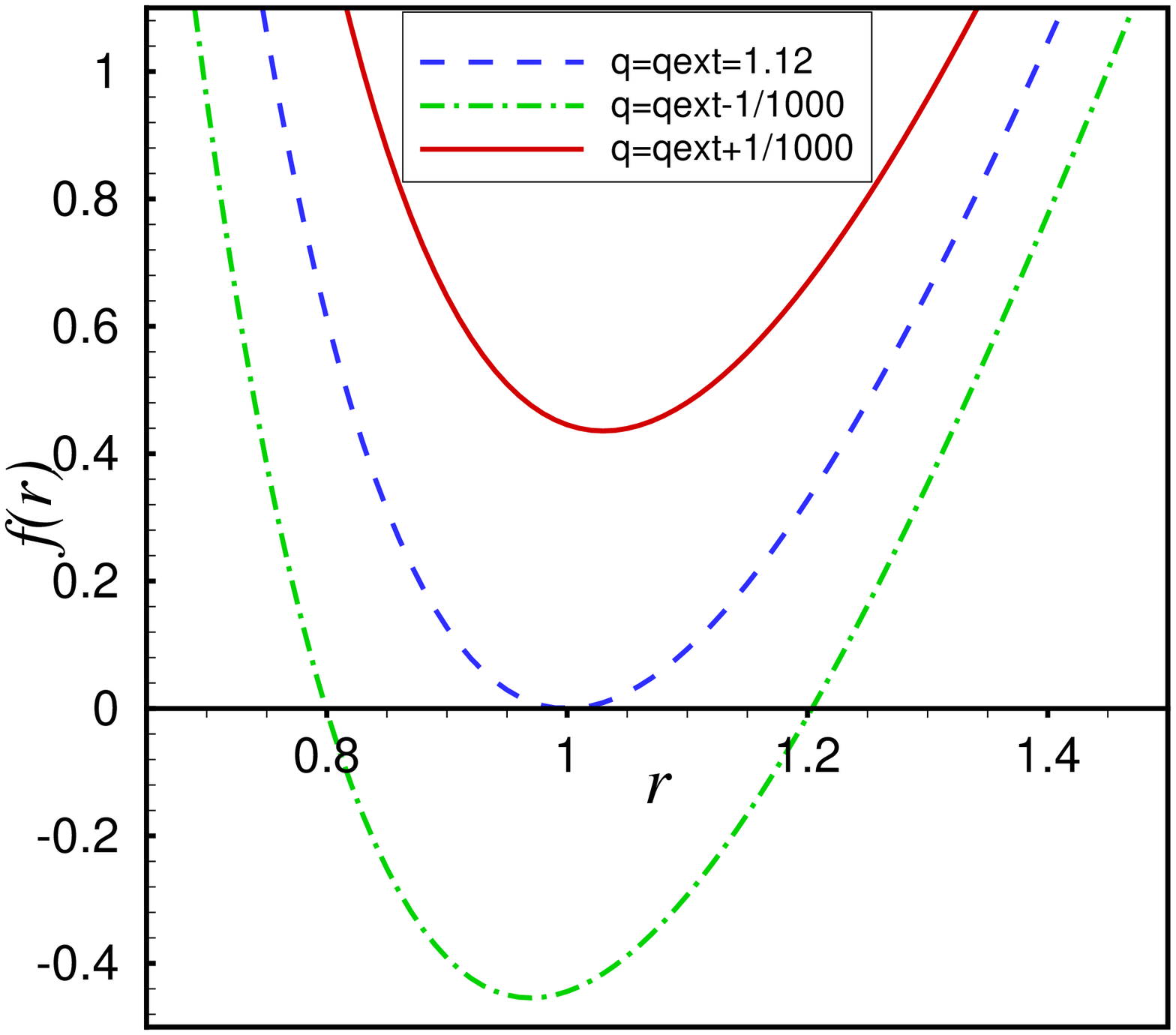}}
\caption{The function $f(r)$ versus $r$ for $n=4$, $\protect\alpha =0.5$, $%
p=2$, $l=b=1$, $k=1$ and $r_{\mathrm{ext}}=1$. }
\label{fig2}
\end{figure}

\begin{figure}[tbp]
\epsfxsize=7cm \centerline{\epsffile{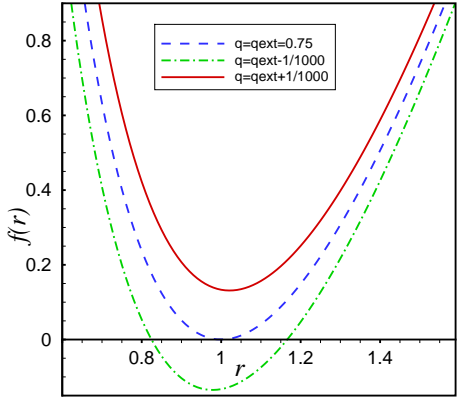}}
\caption{The function $f(r)$ versus $r$ for $n=4$, $\protect\alpha =0.5$, $%
p=2$, $l=b=1$, $k=-1$ and $r_{\mathrm{ext}}=1$.}
\label{fig3}
\end{figure}

\section{Thermodynamics of topological black holes \label{Therm}}

Since discussions on the black holes thermodynamics depend on the mass of
the solutions, we first calculate the mass of the dilaton black holes using
the modified subtraction method of Brown and York (BY) \cite{BY}. In order
to use the modified BY method \cite{modBY}, the metric should be written in
the form
\begin{equation}
ds^{2}=-X({\mathcal{R}})dt^{2}+\frac{d\mathcal{R}^{2}}{Y(\mathcal{R})}+%
\mathcal{R}^{2}d\Omega ^{2}.  \label{Mets}
\end{equation}%
To do this, we perform the following transformation \cite{DB}
\begin{equation*}
\mathcal{R}=rR(r).
\end{equation*}%
It is a matter of calculations to show that the metric (\ref{metric}) may be
written as (\ref{Mets}) with the following $X$ and $Y$:
\begin{eqnarray*}
X(\mathcal{R}) &=&f(r(\mathcal{R})), \\
Y(\mathcal{R)} &=&f(r(\mathcal{R}))\left( \frac{d\mathcal{R}}{dr}\right)
^{2}=\left( \frac{b}{r}\right) ^{2\gamma }\left( 1-\gamma \right) ^{2}f(r(%
\mathcal{R})).
\end{eqnarray*}%
The background metric is chosen to be the metric (\ref{Mets}) with

\begin{eqnarray}
X_{0}(\mathcal{R}) &=&f_{0}(r(\mathcal{R}))=  \notag \\
&&\frac{k\left( n-2\right) {r}^{2\gamma }}{(1-2\gamma )(1-\gamma )\left( {%
\alpha }^{2}+n-2\right) {b}^{2\gamma }}  \notag \\
&&+\frac{2\Lambda {b}^{2\gamma }{r}^{2(1-\gamma )}}{\left( n-1\right)
(1-\gamma )^{2}\left( {\alpha }^{2}-n\right) },  \label{X0}
\end{eqnarray}%
\begin{equation}
Y_{0}(\mathcal{R})=\frac{k\left( 1-\gamma \right) \left( n-2\right) }{%
(1-2\gamma )\left( {\alpha }^{2}+n-2\right) }+\frac{2\Lambda {b}^{4\gamma }{r%
}^{2(1-2\gamma )}}{\left( n-1\right) \left( {\alpha }^{2}-n\right) }.
\label{Y0}
\end{equation}%
To compute the conserved mass of the spacetime, we choose a timelike Killing
vector field $\xi $ on the boundary surface $\mathcal{B}$ of the spacetime (%
\ref{Mets}). Then, the quasilocal conserved mass can be written as
\begin{eqnarray}
M &=&\frac{1}{8\pi }\int_{\mathcal{B}}d^{2}\varphi \sqrt{\sigma }\left\{
\left( K_{ab}-Kh_{ab}\right) \right.  \notag \\
&&\left. -\left( K_{ab}^{0}-K^{0}h_{ab}^{0}\right) \right\} n^{a}\xi ^{b},
\end{eqnarray}%
where $\sigma $ is the determinant of the metric of the boundary $\mathcal{B}
$, $K_{ab}^{0}$ is the extrinsic curvature of the background metric and $%
n^{a}$ is the timelike unit normal vector to the boundary $\mathcal{B}$.
Thus, using the above modified BY formalism, and denoting the volume of
constant curvature hypersurface $h_{ij}dx^{i}dx^{j}$ by$\ \omega _{n-1}$,
one can calculate the mass of the black hole per unit volume ${\omega _{n-1}}
$ as%
\begin{equation}
{M}=\frac{b^{(n-1)\gamma }(n-1)}{16\pi (\alpha ^{2}+1)}m.  \label{Mass}
\end{equation}%
In the following we are going to explore thermodynamics of the topological
dilaton black hole we have just found. The entropy of the topological black
hole typically satisfies the so called area law of the entropy which states
that the entropy of the black hole is a quarter of the event horizon area
\cite{Beck}. This near universal law applies to almost all kinds of black
holes, including dilaton black holes, in Einstein gravity \cite{hunt}. It is
a matter of calculation to show that the entropy of the topological black
hole per unit volume ${\omega _{n-1}}$ is
\begin{equation}
{S}=\frac{b^{(n-1)\gamma }r_{+}^{(n-1)(1-\gamma )}}{4}.  \label{Entropy}
\end{equation}%
Using (\ref{FE3}), the electric charge can be calculated through the Gauss
law

\begin{equation}
Q=\frac{\,{1}}{4\pi }\int e^{-\frac{{4\alpha p\Phi (r)}}{{n-1}}%
}(rR)^{n-1}(-F)^{p-1}F_{\mu \nu }n^{\mu }u^{\nu }d{\Omega },  \label{chdef}
\end{equation}%
where $n^{\mu }$ and $u^{\nu }$ are the unit spacelike and timelike normals
to a sphere of radius $r$ given as

\begin{equation*}
n^{\mu }=\frac{1}{\sqrt{-g_{tt}}}dt=\frac{1}{\sqrt{f(r)}}dt,\text{ \ \ \ \ }%
u^{\nu }=\frac{1}{\sqrt{g_{rr}}}dr=\sqrt{f(r)}dr.
\end{equation*}%
Using (\ref{chdef}), we obtain

\begin{equation}
Q=\frac{\tilde{q}}{4\pi },  \label{Charge}
\end{equation}%
as the charge per unit volume ${\omega _{n-1}}$, where

\begin{equation*}
\tilde{q}=\,{2^{p-1}{q}^{2\,p-1}\,}.
\end{equation*}%
One may note that $\tilde{q}=q$ for $p=1$. The electric potential $U$,
measured at infinity with respect to the horizon, is defined by
\begin{equation}
U=A_{\mu }\chi ^{\mu }\left\vert _{r\rightarrow \infty }-A_{\mu }\chi ^{\mu
}\right\vert _{r=r_{+}},  \label{Pot}
\end{equation}%
where $\chi =C\partial _{t}$ is the null generator of the horizon.
Therefore, using (\ref{At}) the electric potential may be obtained as
\begin{equation}
U=\frac{Cq{b}^{{\frac{\left( 2\,p-n+1\right) \gamma }{\left( 2\,p-1\right) }}%
}}{\Upsilon {r}_{+}^{\Upsilon }}.  \label{Pot1}
\end{equation}

Now, we are in the position to explore the first law of thermodynamics for
the topological dilaton black holes. In order to do this, we obtain the mass
$M$ as a function of extensive quantities $S$, and $Q$. Using the expression
for the charge, the mass and the entropy given in Eqs. (\ref{Mass}), (\ref%
{Entropy}), (\ref{Charge}), and the fact that $f(r_{+})=0$, one can obtain a
Smarr-type formula as

\begin{eqnarray}
M(S,Q) &=&\left( 1+{\alpha }^{2}\right) \left\{ -\frac{{b}^{\alpha
^{2}}{}\Lambda {{(4S)^{\frac{n-\alpha ^{2}}{n-1}}}}}{8{\pi }\left( n-{\alpha
}^{2}\right) }\right.   \notag \\
&&\,\left. +\frac{k\left( n-1\right) \left( n-2\right) {{(4S)^{\frac{\alpha
^{2}+n-2}{n-1}}}}}{16{\pi b}^{\alpha ^{2}}\left( {\alpha }^{2}+n-2\right)
(1-\alpha ^{2})}\right.   \notag \\
&&\left. +\frac{\left( 2p-1\right) ^{2}p(n-1)}{2\Pi \left( {\alpha }%
^{2}\,-2p+n\right) }\left( \frac{\pi {b}^{\alpha ^{2}}}{2^{p-3}}\right) ^{%
\frac{1}{2p-1}}\right.   \notag \\
&&\,\left. \times {Q}^{\frac{2\,p}{2\,p-1}}{{(4S)^{-\frac{\alpha ^{2}-2p+n}{%
\left( 2p-1\right) \left( n-1\right) }}}}\right\} .
\end{eqnarray}%
One may then regard the parameters $S$, and $Q$ as a complete set of
extensive parameters for the mass $M(S,Q)$ and define the intensive
parameters conjugate to $S$ and $Q$. These quantities are the temperature
and the electric potential
\begin{equation}
T=\left( \frac{\partial M}{\partial S}\right) _{Q},\ \ U=\left( \frac{%
\partial M}{\partial Q}\right) _{S},  \label{Dsmar}
\end{equation}%
provided $C$ is chosen as $C=\left( n-1\right) {p}^{2}/\Pi $. It is notable
to mention that $C=1$ in the case of linear Maxwell field \cite{Shey2}.
Calculations show that the intensive quantities calculated by Eq. (\ref%
{Dsmar}) coincide with Eqs. (\ref{Th}) and (\ref{Pot1}) as the temperature
and electric potential. Thus, these thermodynamics quantities satisfy the
first law of thermodynamics
\begin{equation}
dM=TdS+Ud{Q}.
\end{equation}

\begin{figure}[tbp]
\epsfxsize=7cm \centerline{\epsffile{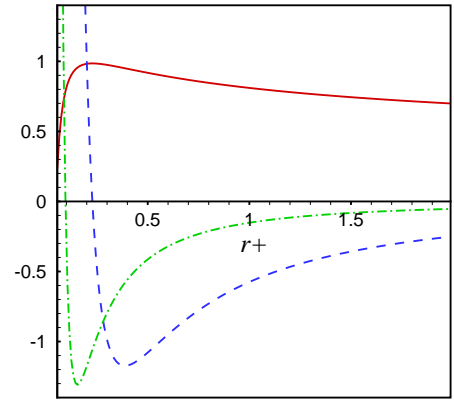}}
\caption{The behavior of $T$ (solid curve), $(\partial ^{2}M/\partial
S^{2})_{Q}$ (dashed curve) and $10^{-2}\mathbf{H}_{S,Q}^{M}$ (dashdot curve)
versus $r_{+}$ for $k=0$ with $l=b=1$, $q=0.4$, $\protect\alpha =1.28$, $n=4$
and $p=2$.}
\label{fig4}
\end{figure}

\begin{figure}[tbp]
\epsfxsize=7cm \centerline{\epsffile{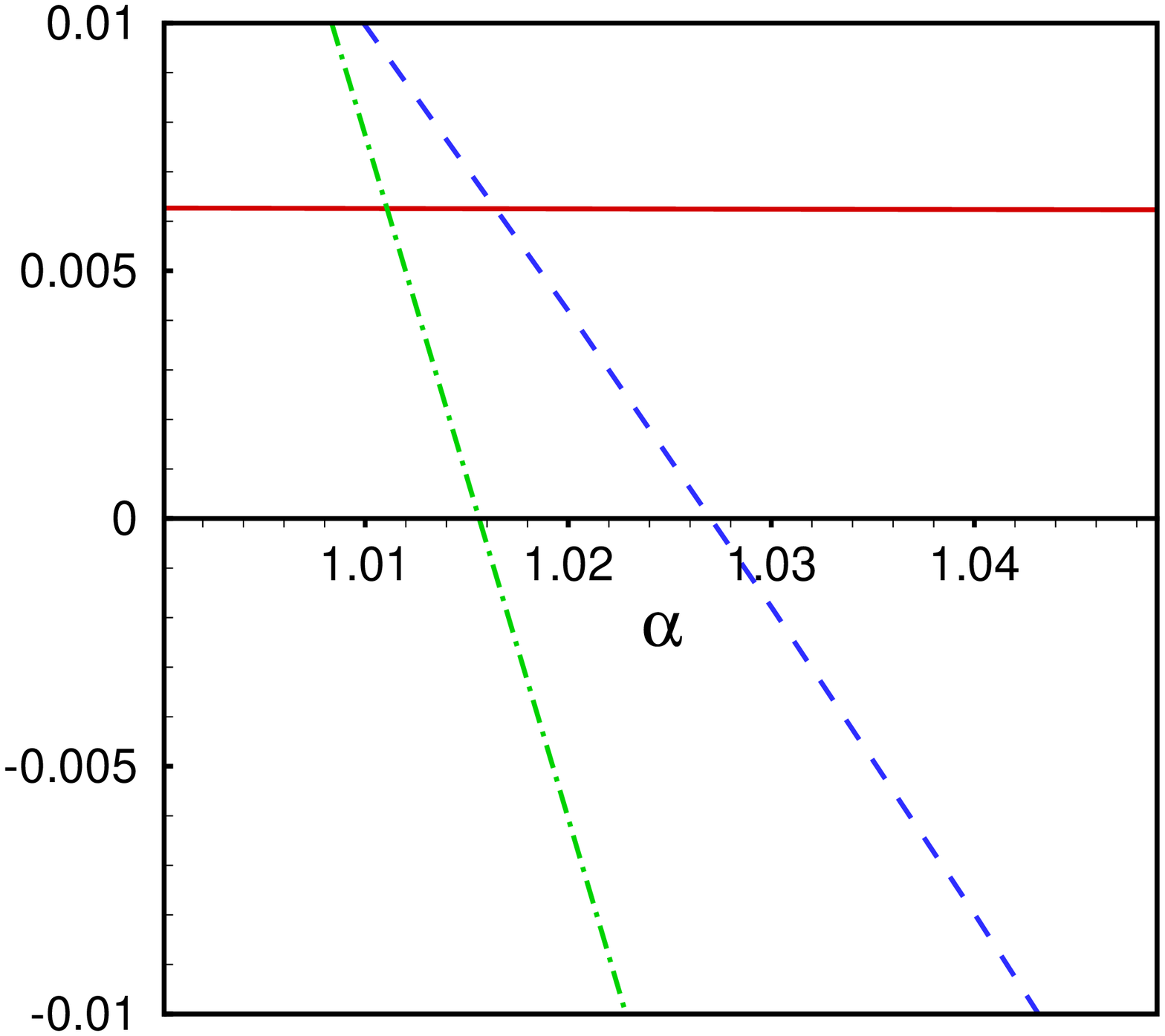}}
\caption{The behavior of $10^{-2}T$ (solid curve), $(\partial ^{2}M/\partial
S^{2})_{Q}$ (dashed curve) and $10^{-1}\mathbf{H}_{S,Q}^{M}$ (dashdot curve)
versus $\protect\alpha $ for $k=0$ with $l=b=1$, $q=0.45$, $r_{+}=2$, $n=4$
and $p=2$.}
\label{fig5}
\end{figure}

\begin{figure}[tbp]
\epsfxsize=7cm \centerline{\epsffile{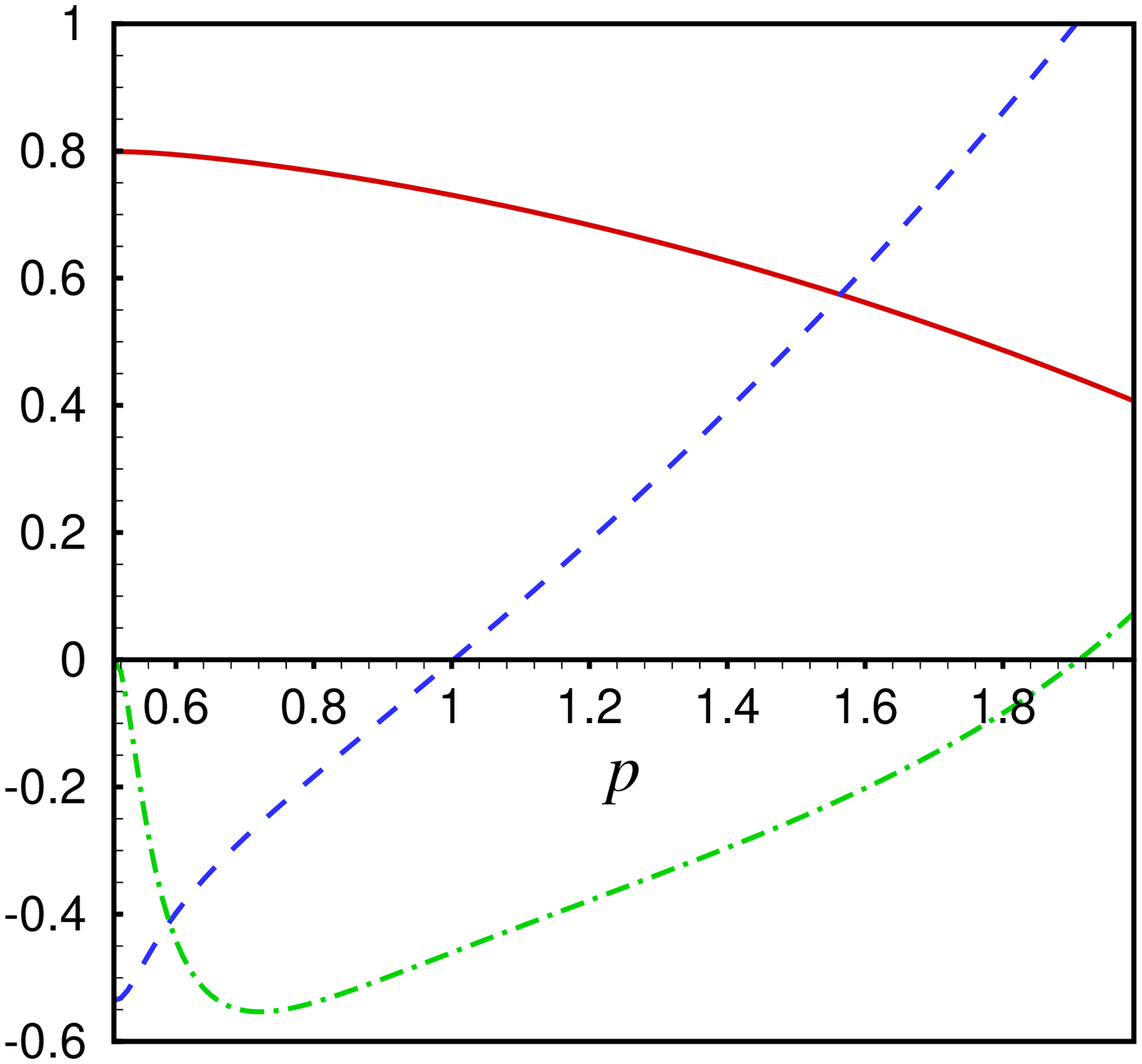}}
\caption{The behavior of $T$ (solid curve), $(\partial ^{2}M/\partial
S^{2})_{Q}$ (dashed curve) and $10^{-1}\mathbf{H}_{S,Q}^{M}$ (dashdot curve)
versus $p$ for $k=0$ with $l=b=1$, $q=0.8$, $r_{+}=1.1$, $\protect\alpha %
=1.25$ and $n=4$.}
\label{fig6}
\end{figure}

\begin{figure}[tbp]
\epsfxsize=7cm \centerline{\epsffile{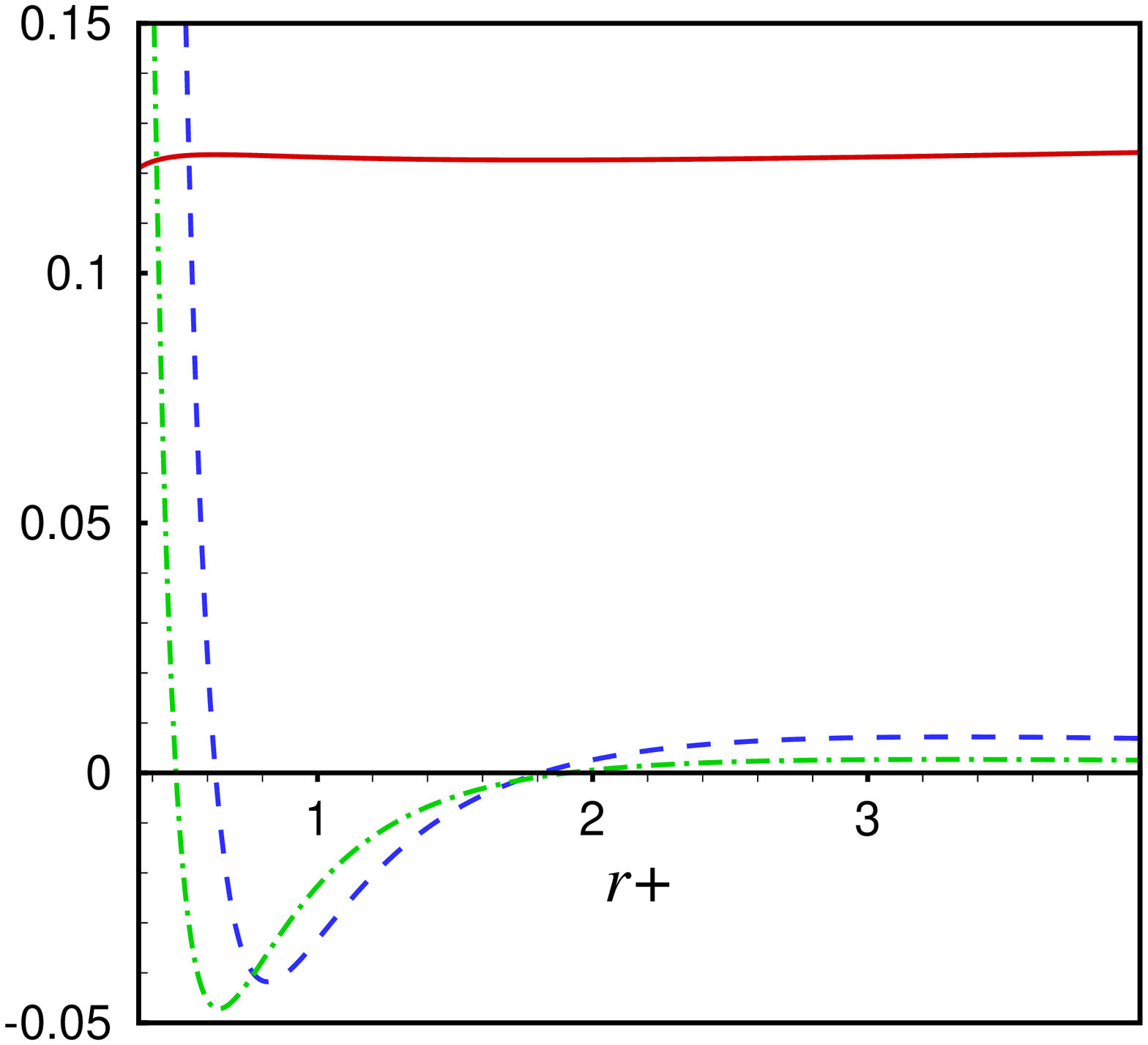}}
\caption{The behavior of $10^{-1}T$ (solid curve), $(\partial ^{2}M/\partial
S^{2})_{Q}$ (dashed curve) and $10^{-2}\mathbf{H}_{S,Q}^{M}$ (dashdot curve)
versus $r_{+}$ for $k=1$ with $l=b=1$, $q=0.4$, $\protect\alpha =0.8$, $n=4$
and $p=2$.}
\label{fig7}
\end{figure}

\begin{figure}[tbp]
\epsfxsize=7cm \centerline{\epsffile{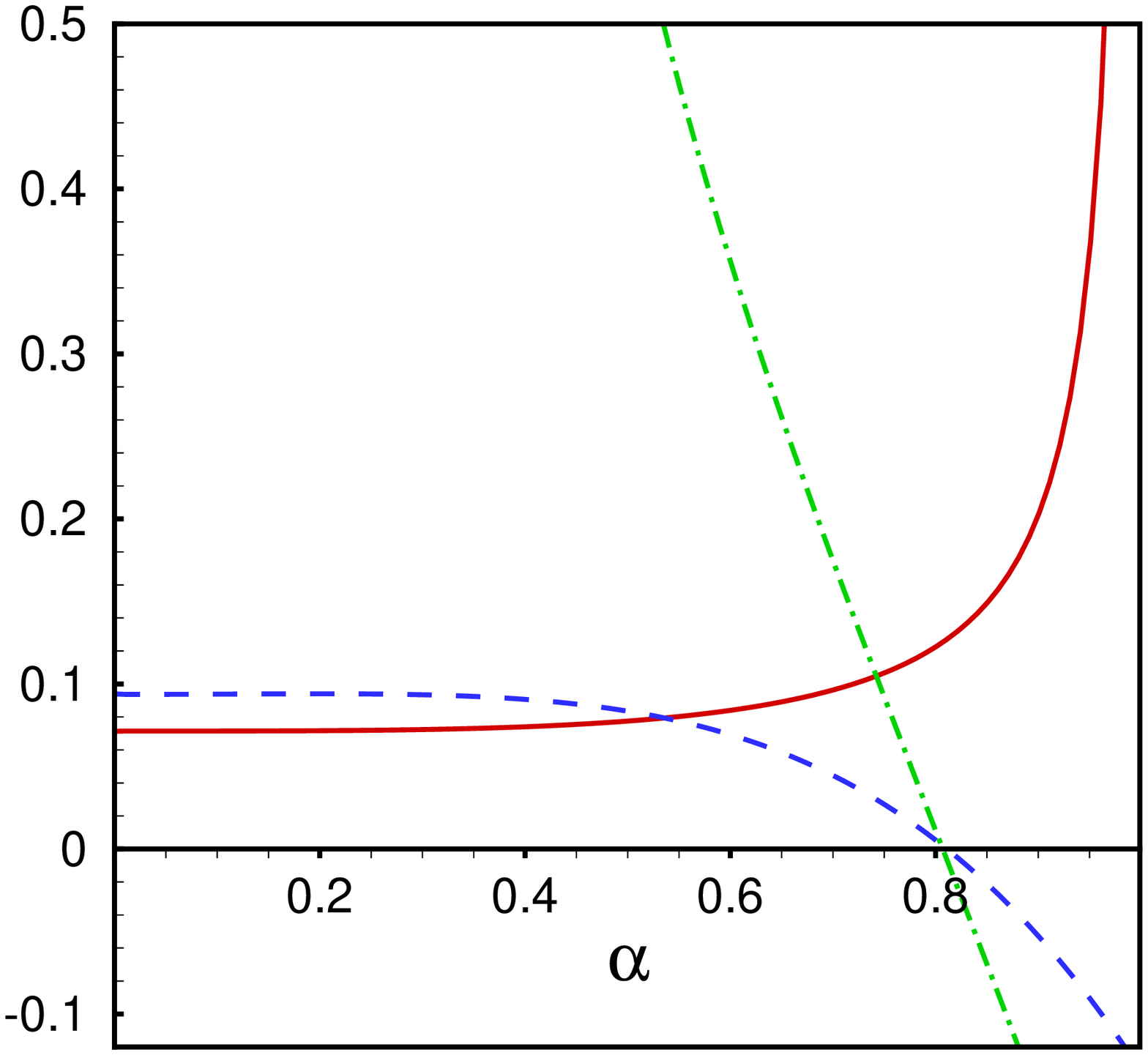}}
\caption{The behavior of $10^{-1}T$ (solid curve), $(\partial ^{2}M/\partial
S^{2})_{Q}$ (dashed curve) and $10^{-1}\mathbf{H}_{S,Q}^{M}$ (dashdot curve)
versus $\protect\alpha $ for $k=1$ with $l=b=1$, $q=0.45$, $r_{+}=2$, $n=4$
and $p=2$.}
\label{fig8}
\end{figure}

\begin{figure}[tbp]
\epsfxsize=7cm \centerline{\epsffile{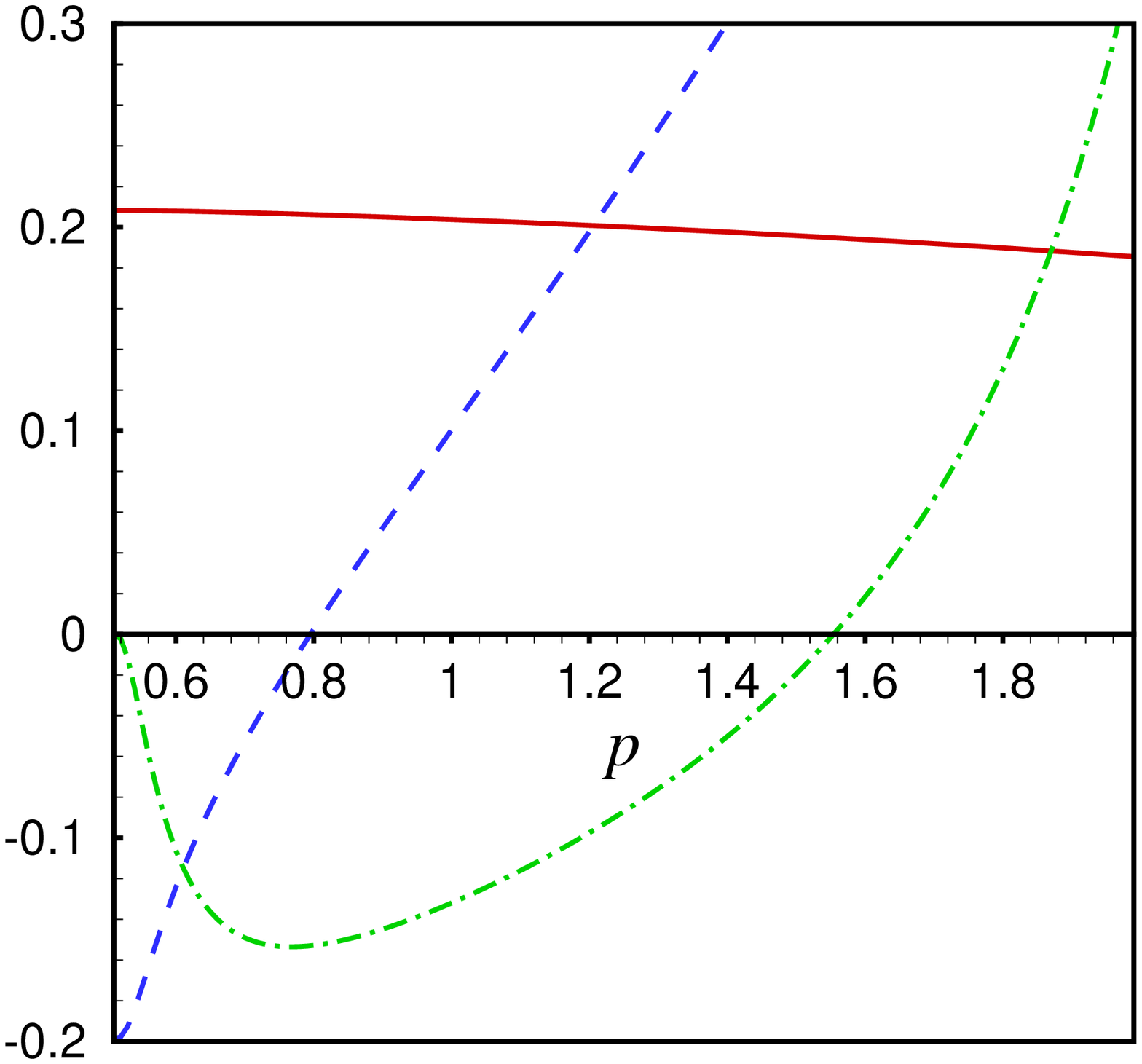}}
\caption{The behavior of $10^{-1}T$ (solid curve), $(\partial ^{2}M/\partial
S^{2})_{Q}$ (dashed curve) and $10^{-1}\mathbf{H}_{S,Q}^{M}$ (dashdot curve)
versus $p$ for $k=1$ with $l=b=1$, $q=0.8$, $r_{+}=1.1$, $\protect\alpha %
=0.9 $ and $n=4$.}
\label{fig9}
\end{figure}

\begin{figure}[tbp]
\epsfxsize=7cm \centerline{\epsffile{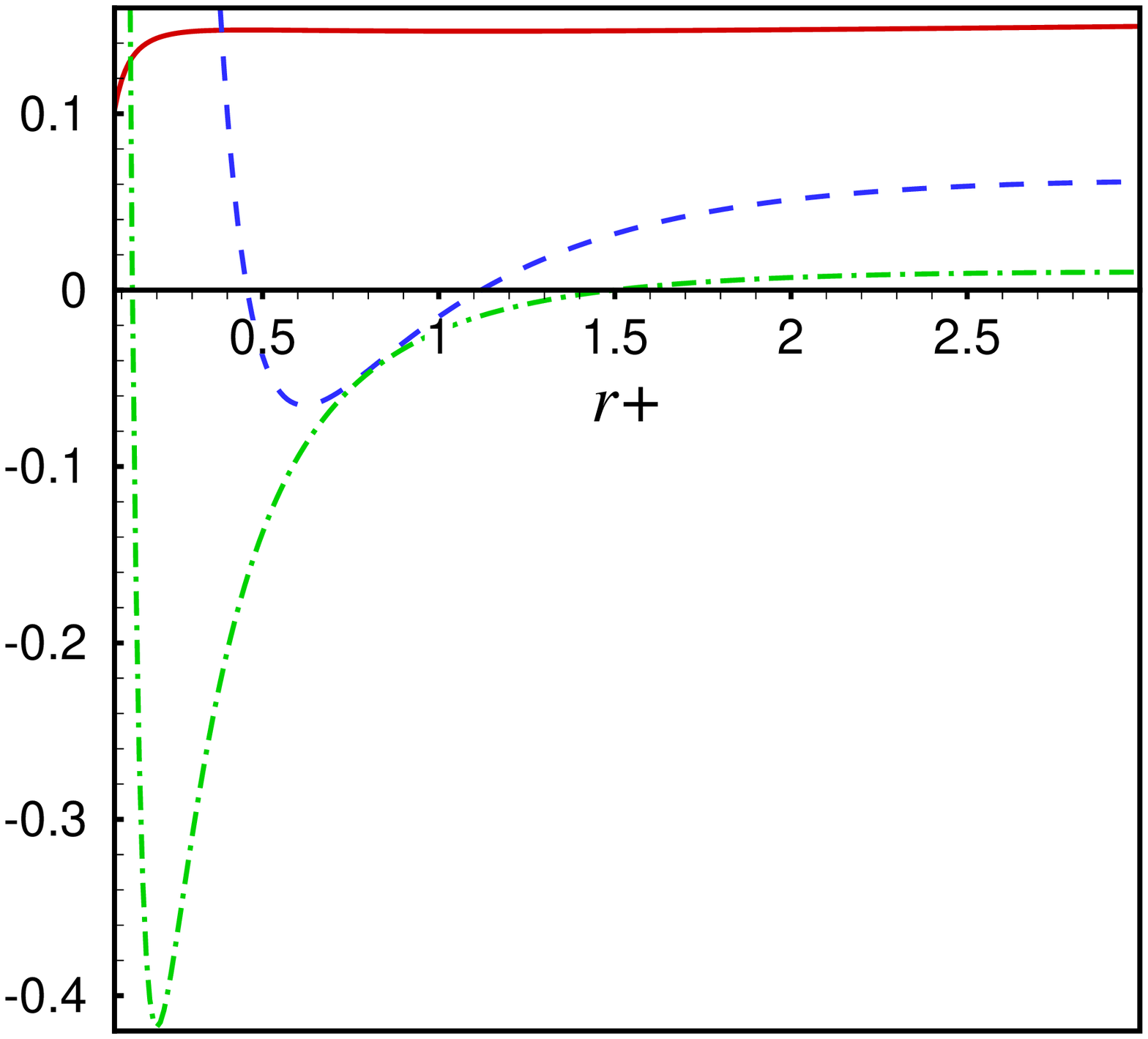}}
\caption{The behavior of $10^{-1}T$ (solid curve), $(\partial ^{2}M/\partial
S^{2})_{Q}$ (dashed curve) and $10^{-2}\mathbf{H}_{S,Q}^{M}$ (dashdot curve)
versus $r_{+}$ for $k=-1$ with $l=b=1$, $q=0.4$, $\protect\alpha =1.28$, $%
n=4 $ and $p=2$.}
\label{fig11}
\end{figure}

\begin{figure}[tbp]
\epsfxsize=7cm \centerline{\epsffile{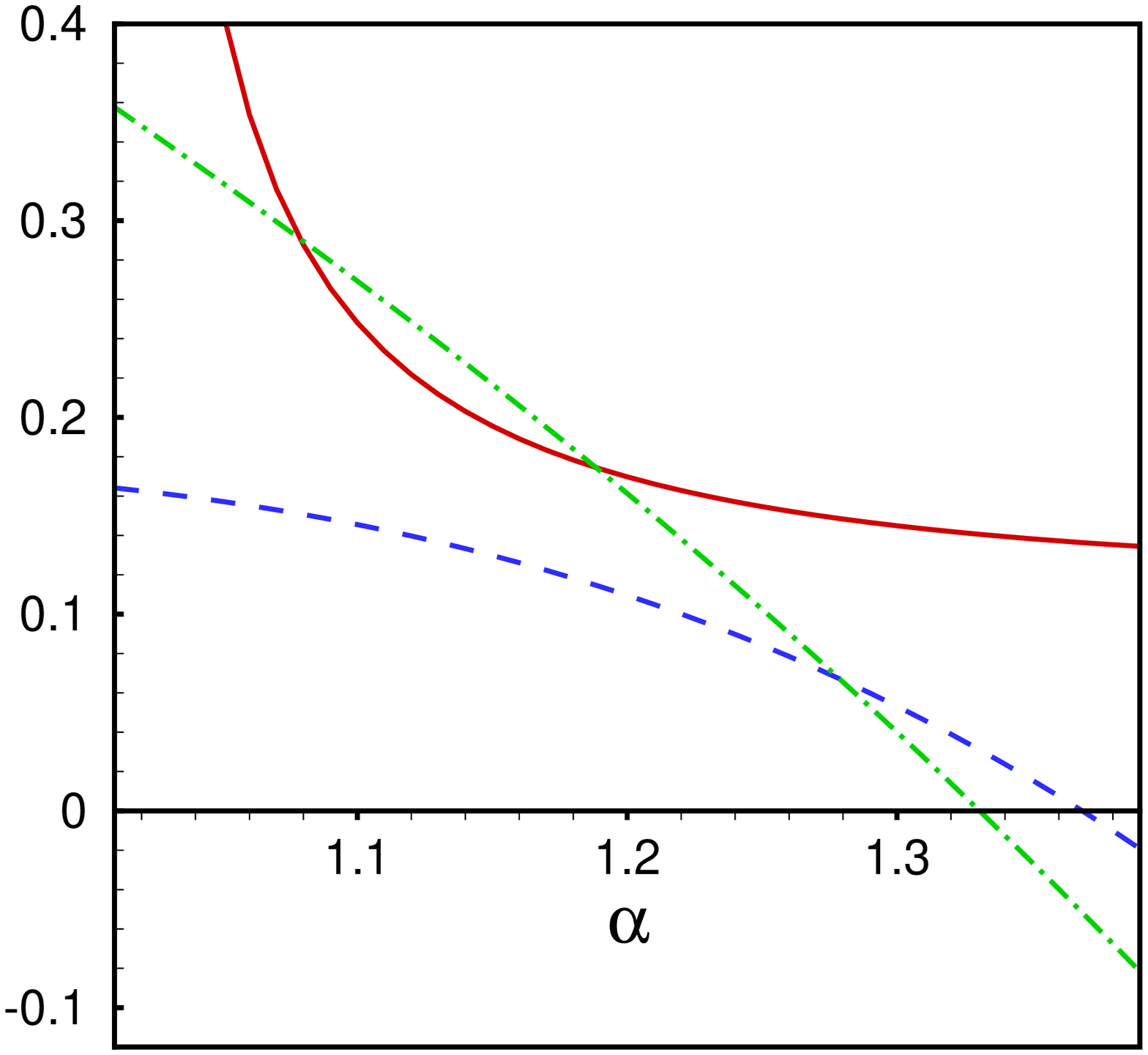}}
\caption{The behavior of $10^{-1}T$ (solid curve), $(\partial ^{2}M/\partial
S^{2})_{Q}$ (dashed curve) and $10^{-1}\mathbf{H}_{S,Q}^{M}$ (dashdot curve)
versus $\protect\alpha $ for $k=-1$ with $l=b=1$, $q=0.45$, $r_{+}=2$, $n=4$
and $p=2$.}
\label{fig10}
\end{figure}

\begin{figure}[tbp]
\epsfxsize=7cm \centerline{\epsffile{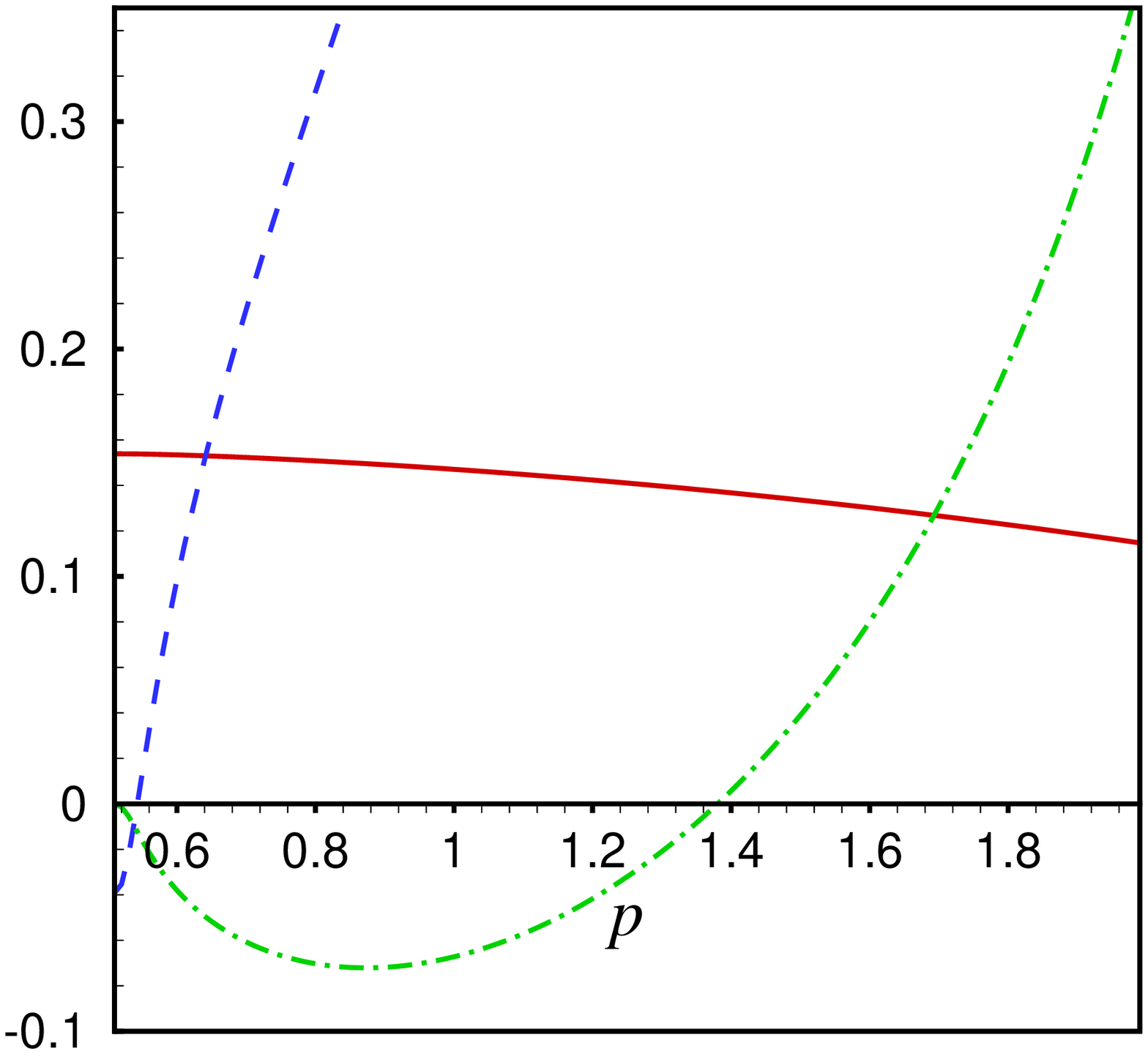}}
\caption{The behavior of $10^{-1}T$ (solid curve), $(\partial ^{2}M/\partial
S^{2})_{Q}$ (dashed curve) and $10^{-1}\mathbf{H}_{S,Q}^{M}$ (dashdot curve)
versus $p$ for $k=-1$ with $l=b=1$, $q=0.8$, $r_{+}=1.1$, $\protect\alpha %
=1.25$, $n=4$.}
\label{fig12}
\end{figure}

\section{Stability in the Canonical and Grand-canonical Ensemble}

\label{Stab}

Finally, we study thermal stability of the topological dilaton black holes.
The stability of a thermodynamic system with respect to small variations of
the thermodynamic coordinates is usually performed by analyzing the behavior
of the entropy $S(M,Q)$ or its Legendre transformation $M(S,Q)$\ around the
equilibrium. The local stability in any ensemble requires that the energy $%
M(S,Q)$ be a convex function of its extensive variable \cite{Cal2,Gub}. The
number of thermodynamic variables depends on the ensemble that is used. In
the canonical ensemble, the charge is a fixed parameter and therefore the
positivity of the heat capacity $C_{v}=T/(\partial ^{2}M/\partial S^{2})_{Q}$
is sufficient to ensure the local stability. Hence, in the ranges where $T$
is positive, the positivity of $(\partial ^{2}M/\partial S^{2})_{Q}$
guarantees the local stability of the solutions. For the spacetime under
consideration we find,

\begin{gather}
\left( \frac{\partial ^{2}M}{\partial S^{2}}\right) _{Q}=\text{ }\frac{%
1+\alpha ^{2}}{\pi (n-1)}\times \text{\ \ \ \ \ \ \ \ \ \ \ \ \ \ \ \ \ \ \
\ \ \ \ \ \ \ \ \ \ \ \ \ \ \ \ \ \ \ \ \ \ \ \ \ \ \ }  \notag \\
\left\{ -\frac{k\left( n-2\right) }{b^{(n+1)\gamma }}\frac{1}{r_{+}^{\delta }%
}+\frac{n(1-\alpha ^{2})}{l^{2}b^{(n-3)\gamma }}\frac{1}{r_{+}^{\vartheta }}%
\right. \text{ \ \ \ \ \ \ \ \ \ \ \ \ \ \ }  \notag \\
\left. +\frac{2^{p}p(2p(n-2)+1+\alpha ^{2}){q}^{2\,p}}{{b}^{\frac{\gamma
\lbrack 2(2n-3)p-n+1]}{2p-1}}\Pi }\frac{1}{r_{+}^{\eta }}\right\} ,
\label{ddms}
\end{gather}%
where $\delta =(n-\alpha ^{2})/(1+\alpha ^{2})$, $\eta =[\alpha
^{2}+2p(2n-3)-n+2]/[(2p-1)(1+\alpha ^{2})]$ and $\vartheta =(\alpha
^{2}+n-2)/(1+\alpha ^{2})$. In grand-canonical ensemble $Q$ is no longer
fixed. In our case, the mass is a function of entropy and charge and
therefore the system is locally stable provided $\mathbf{H}_{SQ}^{M}=\left[
\partial ^{2}M/\partial S\partial Q\right] >0$, where the determinant of
Hessian matrix can be calculated as

\begin{eqnarray}
\mathbf{H}_{S,Q}^{M} &=&\frac{2^{\frac{-2p^{2}+13p-9}{2p-1}}\left( {\alpha }%
^{2}+1\right) ^{2}p^{2}{b}^{\frac{\alpha ^{2}}{2p-1}}}{\left( {\alpha }%
^{2}\,-2p+n\right) \Pi {{q}^{2p-2}}}  \notag \\
&&\times \left[ \frac{16(2p-1)p(2p-1-\alpha ^{2})q^{2p}}{\Pi }\right.  \notag
\\
&&\times \left( 2^{2p^{2}-3p-4}{b}^{\alpha ^{2}}\right) ^{\frac{1}{2p-1}}{{%
(b^{\gamma }r_{+}^{(1-\gamma )})^{-\frac{2(\alpha ^{2}+2p(n-2)+1)}{\left(
2p-1\right) }}}}  \notag \\
&&-\frac{k\left( n-2\right) }{{b}^{\alpha ^{2}}}{{(b^{\gamma
}r_{+}^{(1-\gamma )})^{-\frac{\alpha ^{2}-2p+n+\left( 2p-1\right) (n-\alpha
^{2})}{\left( 2p-1\right) }}}}  \notag \\
&&\left. +\frac{n{b}^{\alpha ^{2}}{}}{{l}^{2}}(1-\alpha ^{2}){{(b^{\gamma
}r_{+}^{(1-\gamma )})^{-\frac{2(p(\alpha ^{2}+n-3)+1)}{\left( 2p-1\right) }}}%
}\right] .  \label{Hes}
\end{eqnarray}%
Here, we discuss the stability of the black hole solutions for different
values of $k$ separately.

(i) $k=0$: In this case, one can see that (\ref{ddms}) is always
positive for $\alpha \leq 1$, and therefore the black holes with
$k=0$ and $\alpha \leq 1$ are thermally stable in the canonical
ensemble. However, there may
exist unstable black holes in the grand-canonical ensemble for the range $%
2p-1<\alpha ^{2}\leq 1$. Of course, one should note that the black holes in
both the canonical and grand-canonical ensembles only exist provided $q<q_{%
\mathrm{ext}}$. For $\alpha >1$, only small black holes with event horizon
radius less than $r_{+}^{\mathrm{\max }}$ is stable, where the value of $%
r_{+}^{\mathrm{\max }}$ is smaller in grand-canonical ensemble as one may
see in Fig. \ref{fig4}. Figure \ref{fig5} shows the effects of $\alpha $ on
the stability of the solutions in both canonical and grand canonical
ensembles. As one increases the coupling constant $\alpha $, there is an $%
\alpha _{\max }$ that for $\alpha <\alpha _{\max }$, black holes are stable.
The value of $\alpha _{\max }$ depends on the ensemble and it is larger in
the canonical one. Figure \ref{fig6} shows the effects of $1/2<p<n/2$ on the
stability of the solutions. This figure shows that there is a $p_{\min }$
that for $p>p_{\min }$, black holes are stable. Again $p_{\min }$ is
ensemble dependent and it is larger in grand-canonical ensemble. For $%
n/2<p<n-1$ where $\alpha $ has a $p$-dependent lower limit, numerical
analysis shows that there exists a minimum value $p_{\min }$ for which black
holes are stable provided $p>p_{\min }$.

(ii) $k=1$: In this case from figure \ref{fig7} we see that the Hawking-Page
phase transition occurs between small and large black holes. Choosing $%
q<\left( q_{ext}\right) _{\alpha =0}$, then one ensure that $T>0$ for the
allowed region $\alpha <1$. Again, there is a maximum value for $\alpha $
such that black holes are stable for $\alpha <\alpha _{\max }$ (See Fig. \ref%
{fig8}). The effects of $p$ in the range $1/2<p<n/2$ on the
stability both in canonical and grand canonical ensembles are
shown in Fig. \ref{fig9}. One can see that for $p>p_{\min }$,
these solutions represent stable black holes.
Numerical calculations show that for $n/2<p<n-1$ where $\alpha $ has a $p$%
-dependent lower limit, there is also a minimum value $p_{\min }$ that for
values greater than $p_{\min }$, the black holes are stable.

(iii) $k=-1$: As in the case of $k=0$, the stability of black
holes in both canonical and grand-canonical ensembles should be
investigated separately for $\alpha <1$ and $\alpha >1$ cases. As
one can see from Eq. (\ref{ddms}), $\left( \partial ^{2}M/\partial
S^{2}\right) _{Q}$ is positive for $\alpha <1 $. Therefore, for
$q<q_{\mathrm{ext}}$, which black holes exist, they are stable.
However, in the grand-canonical ensemble, black holes may be
unstable in the range $2p-1<\alpha ^{2}\leq 1$. For $\alpha >1$,
as one can
see in Fig. (\ref{fig10}), the black hole solutions are stable provided $%
\alpha <\alpha _{\max }$. Of course, the value of $\alpha _{\max }$ depends
on the ensemble. Figure (\ref{fig11}) shows a Hawking-Page phase transition
between small and large black holes.

In order to investigate the effect of $p$ on the stability of the solutions,
we plot both $(\partial ^{2}M/\partial S^{2})_{Q}$ and the determinant of
Hessian matrix versus $p$ in the range $1/2<p<n/2$. This is plotted in Fig. (%
\ref{fig12}), which shows that the black hole solutions are stable provided $%
p>p_{\min }$. Numerical calculations also show that for $n/2<p<n-1$ where $%
\alpha $ has a $p$-dependent lower limit, black holes are stable.
\section{Dynamical Stability of 4-dimensional Black Hole Solutions}
Besides thermal stability, it is worthwhile to study the dynamical
stability of solutions under perturbations. Since study of
dynamical stability for higher-dimensional topological solutions
are difficult in general, we study the case of $4$-dimensional
black holes. Regge and Wheeler showed that~in  $4$-dimensional
static and spherically symmetric background, perturbations can be
decomposed into odd-and even-parity sectors according to their
transformation properties under a two-dimensional rotation
\cite{regwhe}.
Perturbations also can be decomposed into sum of spherical harmonics $%
Y_{m}^{\ell }$. In the Regge-Wheeler formalism, stability is
investigated by studying the behaviour of perturbation modes.

Using pointed out formalism, it is shown that in the framework of
scalar-tensor gravity models with general form of the action as\cite{scteodd}%
\begin{eqnarray}
S&=&\int d^{4}x\sqrt{-g}\left[ G(\Phi ,Z)\mathcal{R}+K(\Phi
,Z)\right. \nonumber
\\
&& \left.+G_{Z}\left[ (\square \Phi )^{2}-(\nabla _{\mu }\nabla
_{\nu }\Phi )^{2}\right] \right] , \label{stgm}
\end{eqnarray}%
where $G$ and $K$ are arbitrary functions of $\Phi $ and
$Z=-(\nabla \Phi )^{2}/2$ and $G_{Z}=\partial G/\partial Z$, there
are dynamically stable solutions under odd-type perturbations
provided
\begin{equation}
\mathcal{F}:=2G>0,\text{ \ \ \ \ }\mathcal{G}:=2G-4ZG_{Z}>0,
\label{con1}
\end{equation}%
when the single mode propagets radially with the squared speed of $c_{r}^{2}=%
\mathcal{G}/\mathcal{F}$. In the case of uncharged solutions,
(\ref{stgm}) is match with our action (\ref{Act}) provided $G=1$
and $K=Z-V(\Phi )$. It is obvious from (\ref{con1}) that in this
case our solutions are dynamically stable. Under even-type
perturbations, there is again a single mode in our case that
propagates with the same radial speed as odd-type perturbations
\cite{scteeven}. In this case we have stable solutions provided
$\ell \geq 2$.

Dynamical stability of non-linear electrodynamics (NED) sources in
general relativity is studied in \cite{nedsta}. In the case of a
NED Lagrangian $\mathcal{L}(\hat{F})$ where $\hat{F}=1/4F$, the
corresponding
Hamiltonian can be defined as $\mathcal{H}\equiv 2\mathcal{L}_{\hat{F}}\hat{F%
}-\mathcal{L}$. It is also convenient to study the stability using
so-called $P$ frame where $P=\mathcal{L}_{\hat{F}}^{2}\hat{F}$.
For odd-type perturbations, there are stable solutions provided
$\mathcal{H}_{P}$ vanishes no where outside the horizon while for
even-type ones, we encounter instability provided
$\mathcal{H}_{xx}>0$\ where $x=\sqrt{-2Q^{2}\,P}$. In our case
with $\alpha =0$, one can calculate

\begin{equation}
\mathcal{H}_{P}=\frac{1}{p}\left( \frac{-4P}{p^{2}}\right) ^{\frac{1-p}{2p-1}%
},  \label{Hp}
\end{equation}%
where $P=-p^{2}\left( 2F_{tr}^{2}\right) ^{2p-1}/4$. Obviously $\mathcal{H}%
_{P}$ vanishes no where outside the horizon and therefore under
odd-type perturbations we have stable solutions. Since

\begin{equation}
\mathcal{H}_{xx}=\frac{x\mathcal{H}_{P}}{Q^{4}}\left[ 1+\frac{\sqrt{2Q^{2}}%
\left( p-1\right) }{2p-1}\left( -P\right) ^{{3}/{2}}\right] ,
\end{equation}%
one encounters dynamically unstable solutions for $p\geq 1$.
\section{Summary and Conclusions}

To sum up, we generalized the investigations on the power-law Maxwell field
to dilaton gravity. We first proposed the suitable Lagrangian in the
Einstein-dilaton gravity in the present of power-law Maxwell Lagrangian
which is coupled to the dilaton field as $[-\exp (-4\alpha \Phi
/(n-1))F_{\mu \nu }F^{\mu \nu }]^{p}$. Then, we constructed a new class of $%
(n+1)$-dimensional $(n\geq 3)$ topological black hole solutions of
this theory in the presence of Liouville-type potentials for the
dilaton field. In contrast to the topological black holes of EMd
gravity \cite{Shey2} which exits for the Liouville-type potentials
with two terms, here we found that the solutions exist provided we
assumes three Liouville-type potentials for the dilaton field. In
the limiting case where $p=1$ one of the Liouville potential
vanishes. Due to the presence of the dilton field, the obtained
solutions are neither asymptotically flat nor (A)dS. Besides, for
the cases
of $k=\pm 1$ the solutions do not exist for the string case where $\alpha =1$%
. When $p=1$, all results of toplogical black holes of EMd gravity are
recovered \cite{Shey2}.

The facts that (i) the gauge potential of electromagnetic field
$A_{t}$ is finite, (ii) the dilaton potential $V(\Phi )$ has
finite lower limit, and (iii)  the terms contain mass and charge
in the metric function (\ref{f}) should be disappeared in the
spacial infinity, imply that the parameters $p$ and $\alpha $
should be restricted as follows. For $1/2<p<n/2$, we should have
$\alpha ^{2}<n-2$, while for $n/2<p<n-1$, we should have
$2p-n<\alpha ^{2}<n-2$. Requiring the fact that our solutions
should be positive in the
spacial infinity, leads to another restriction on $\alpha $ in the case of $%
k=1$, namely $\alpha <1$. Our solutions are well-defined in the
permitted ranges of $p$ and $\alpha $, while in the case of linear
Maxwell field the solutions are ill-defied for $\alpha =\sqrt{n}$.
We showed that our solution can not represent black holes with
single event horizon. However, they can represent black holes with
two horizon, extreme black holes and naked singularity depending
on the model parameters. We also calculated the charge, mass,
temperature, entropy and electric potential of the topological
dilaton black holes and found that the first law of thermodynamics
is satisfied on the black hole horizon. By calculating the
Smarr-type formula, $M(S,Q)$, we analyzed thermal stability of the
solutions in both canonical and grand-canonical ensembles. We
showed that for $\alpha
<1$, there are stable black holes in the cases of $k=0,-1$ provided $q<q_{%
\mathrm{ext}}$ in canonical ensemble whereas in grand-canonical ensemble
black holes may be unstable in the range of $2p-1<\alpha ^{2}\leq 1$. In the
cases of (i) $k=0,-1$, and $\alpha >1$ and (ii) $k=1$ and $\alpha <1$, there
is a maximum value for the dilaton coupling constant $\alpha _{\mathrm{max}}$
for which the obtained solutions are thermally unstable provided $\alpha
>\alpha _{\mathrm{max}}$. For $k=-1$, $\alpha >1$ and $k=1$, $\alpha <1$,
there is a Hawking-Page phase transition between small and large
black holes, while for $k=0$, $\alpha >1$, large black holes are
unstable. This fact can be understood from figs. \ref{fig4},
\ref{fig7} and \ref{fig11}. We also found that there is a $p_{\min
}$ for which we have stable black holes provided $p>p_{\min }$.
Finally, we discussed the dynamical stability of the obtained
solutions in the absence and presence of the non-linear
electromagnetic source, separately.

Note that the $(n+1)$-dimensional charged topological dilaton black holes
obtained here are static. Thus, it would be interesting if one could
construct charged rotating black holes/branes in $(n+1)$ dimensions in the
presence of dilaton and power-law Maxwell field. These issues are now under
investigation and the results will be appeared elsewhere.

\acknowledgments{We thank referee for constructive comments which
helped us improve the paper significantly. We also thank from the
Research Council of Shiraz University. This work has been
supported financially by Research Institute for Astronomy \&
Astrophysics of Maragha (RIAAM), Iran.}

\end{document}